\documentclass[twocolumn,10pt]{IEEEtran}
\usepackage{ifpdf, flushend,subfigure}

\ifCLASSINFOpdf
  \usepackage[pdftex]{graphicx}
  \graphicspath{{../pdf/}{../jpeg/}}
  \DeclareGraphicsExtensions{.pdf,.jpeg,.png}
\else
  \usepackage[dvips]{graphicx}
  \graphicspath{{../eps/}}
  \DeclareGraphicsExtensions{.eps}
\fi
\usepackage[cmex10]{amsmath}
\usepackage {amssymb}
\usepackage{algorithmic}
\usepackage{array}
\usepackage{mdwmath}
\usepackage{mdwtab}
\usepackage{eqparbox}
\usepackage{algorithm}
\usepackage{algorithmic}

\newcommand{\argmax}{\operatornamewithlimits{argmax}}
\newcommand{\beq}{\begin{equation}}
\newcommand{\eeq}{\end{equation}}
\newcommand{\beqn}{\begin{eqnarray}}
\newcommand{\eeqn}{\end{eqnarray}}
\newcommand{\beqno}{\begin{eqnarray*}}
\newcommand{\eeqno}{\end{eqnarray*}}
\newcommand{\bma}{\begin{displaymath}}
\newcommand{\ema}{\end{displaymath}}
\newcommand{\bnu}{\begin{enumerate}}
\newcommand{\enu}{\end{enumerate}}
\newcommand{\bce}{\begin{center}}
\newcommand{\ece}{\end{center}}
\newcommand{\btb}{\begin{tabular}}
\newcommand{\etb}{\end{tabular}}

\begin{document}
%
\title{Channel Assignment for Throughput Maximization in Cognitive Radio Networks}

\author{\IEEEauthorblockN{Le Thanh Tan and Long Bao Le}  
\thanks{The authors are with INRS-EMT, University of Quebec,  Montr\'{e}al, Qu\'{e}bec, Canada. 
Emails: \{lethanh,long.le\}@emt.inrs.ca. }}

\maketitle

\begin{abstract}
In this paper, we consider the channel allocation problem for throughput maximization in
cognitive radio networks with hardware-constrained secondary users. Specifically,
we assume that secondary users exploit spectrum holes on a set of channels
where each secondary user can use at most one available channel for communication. 
We develop two channel assignment algorithms that can efficiently utilize spectrum
opportunities on these channels. In the first algorithm, secondary users are 
assigned distinct sets of channels. We show that this algorithm achieves the maximum throughput limit
if the number of channels is sufficiently large. In addition, we propose an overlapped
channel assignment algorithm, that can improve the throughput performance compared to
the non-overlapped channel assignment algorithm. In addition, we design a distributed
MAC protocol for access contention resolution and  integrate the derived MAC protocol overhead into
the second channel assignment algorithm. Finally, numerical results are presented
to validate the theoretical results and illustrate the performance gain due to the overlapped 
channel assignment algorithm.

\end{abstract}


\section{Introduction}

Emerging broadband wireless applications have been demanding unprecedented increase in radio spectrum resources.
As a result, we have been facing a serious spectrum shortage problem. However, several recent
measurements reveal very low spectrum utilization in most useful frequency bands \cite{Zhao07}.
Cognitive radio technology is a promising technology that can fundamentally improves the
spectrum utilization of licensed frequency bands through secondary spectrum access.
However, transmissions from primary users should be satisfactorily protected from 
secondary spectrum access due to their strictly higher access priority. Protection
of primary communications can be achieved through interference avoidance or interference
control approach (i.e., spectrum overlay or spectrum underlay) \cite{Zhao07}.

For the interference control approach, transmission powers of secondary users
should be carefully controlled so that the aggregated interference they create
at primary receivers does not severely affect ongoing primary communications \cite{Le08}.
In most practical scenarios where direct coordination between primary and second users
is not possible and/or when distributed communications strategies are desired, it would be very
difficult to maintain these interference constraints. The interference avoidance approach
instead protects primary transmissions by requesting secondary users to perform spectrum
sensing to discover spectrum holes over which they can transmit data. Developing efficient 
spectrum sensing and access mechanisms have been very active research topics in the last several years 
\cite{Yu09}-\cite{Su08}. This paper focuses on developing efficient channel assignment  
algorithms for spectrum sharing in a cognitive radio network with hardware-constrained secondary nodes. 

In particular, we consider the scenario where each secondary user can exploit only one
available channel for communications. This can be the case where secondary users' transceivers
are equipped with only one radio with a narrow-band RF front end \cite{So04}. In addition, it is assumed that white spaces are 
so dynamic that it is not affordable for each secondary user to sense all channels to discover
available ones and/or to exchange sensing results with one another. Under this setting, we
are interested in determining a set of channels allocated for each secondary user in advance
so that  maximum network throughput can be achieved in a distributed manner. To the best of
our knowledge, this important problem has not been considered before.

Because the underlying problem is NP-hard, we develop two greedy non-overlapped and overlapped
channel assignment algorithms, which can work very efficiently. In addition, we design and analyze a 
distributed MAC protocol which is integrated into the overlapped channel assignment algorithm.
We demonstrate through numerical studies that if the number of channels is large then the proposed non-overlapped
channel assignment works efficiently. In addition, the overlapped channel assignment algorithm
can achieve noticeable network throughput improvement compared to the non-overlapped counterpart if the number of channels is small or moderate.

The remaining of this paper is organized as follows. Section ~\ref{SystemModel} describes the system model and problem formulation. 
We present a non-overlapped channel assignment algorithm and describe its performance in Section III.
Development of overlapped channel assignment and the corresponding MAC protocol is considered in Section IV. Section ~\ref{Results} 
demonstrates numerical results followed by concluding remarks in Section ~\ref{conclusion}.
 
\section{System Model and Problem Formulation}
\label{SystemModel}


\subsection{System Model}
\label{System}

We consider the collocated cognitive radio network in which $M$ secondary users exploit spectrum opportunities in  $N$ channels. We assume that each secondary user can use at most one channel for his/her data transmission.
In addition, time is divided fixed-size cycle where secondary users perform sensing on assigned channels
at the beginning of each cycle to explore available channels for communications.
 We assume that sensing time is negligible compared to the cycle time and there is no sensing error.
It is assumed that secondary users transmit at a constant rate which is normalized to 1 for throughput calculation purposes.

We consider two different channel assignment schemes. In the first scheme, secondary users
are assigned distinct sets of channels. This channel assignment scheme simplifies the spectrum sharing
design because secondary users do not compete for the same available channels. However, it overlooks
the potential diversity gain of the spectrum sharing problem.
In the second scheme, we allow secondary users to sense and operate on overlapped channels. 
When one particular channel is exploited by several secondary users, it is assumed that
a Medium Access Control (MAC) protocol is employed to resolve the channel contention.

\subsection{Problem Formulation}
\label{ProbForm}

We are interested in performing channel assignment to maximize the system throughput. Let $T_i$
denote the throughput achieved by secondary user $i$. 
Let $x_{ij}$ describe the channel assignment decision where $x_{ij}=1$ if channel $j$ is assigned to secondary 
user $i$ and $x_{ij}=0$, otherwise. 
Then, the throughput maximization
problem can be formally written as follows:
\vspace{0.0cm}
\beqn
\label{Tput}
&& \mathop {\max }\limits_\textbf{x} \sum\limits_{i = 1}^M {{T_i}}.  \label{obj1} 
\eeqn
For non-overlapped channel assignments, we have following constraints
\beqn
 \sum\limits_{i = 1}^M {{x_{ij}} = 1}, \quad  \mbox{for\:all}\: j  \label{con1}.
\eeqn
We can derive the throughput achieved by secondary user $i$ for non-overlapped channel assignment as follows.
Let $S_i$ be the set of channels assigned to secondary user $i$. Let $p_{ij}$ be the probability that 
channel $j$ is available at secondary user $i$. For simplicity, we assume that $p_{ij}$ are independent from one another.
This assumption holds when each secondary user impact different set of primary users on each channel. This can indeed be the case because 
spectrum holes depend on space. Note, however, that 
this assumption can be relaxed if the dependence structure of these probabilities is available. Under this assumption, $T_i$ can be calculated as
\beq \label{tput1}
T_i = 1 - \prod_{j \in \mathcal{S}_i} \overline{p}_{ij} = 1 - \prod\limits_{j = 1}^N {{{\left( {{{\bar p}_{ij}}} \right)}^{{x_{ij}}}}}
\eeq
where $\overline{p}_{ij} = 1 - {p}_{ij}$ is the probability that channel $j$ is not available for 
secondary user $i$. In fact,   $1 - \prod_{j \in \mathcal{S}_i} \overline{p}_{ij}$ is the probability that
there is at least one channel available for secondary user $i$. Because each secondary user can use
at most one available channel, its maximum throughput is 1. 
In the overlapped channel assignment scheme, constraints in (\ref{con1}) are not needed. 
From this calculation, it can be observed that the optimization problem (\ref{obj1})-(\ref{con1})
is a non-linear integer program, which is a NP-hard problem. Given the large computational complexity
required the considered problem, we will develop sub-optimal and low-complexity channel assignment
algorithms in the following.

\section{Non-overlapped Channel Assignment Algorithm}
\label{nonover}

We develop a low-complexity  algorithm for non-overlapped channel assignment in this section. 
Recall that $\mathcal{S}_i$ is the set of channels assigned for secondary
user $i$. In the non-overlapped channel assignment scheme, we have $\mathcal{S}_i \cap \mathcal{S}_j = \emptyset, \: i \neq j$. The greedy
channel assignment algorithm iteratively allocates channels to secondary users that achieves maximum increase
in the throughput. Detailed description of the proposed algorithm is presented in Algorithm 1.
In each channel allocation iteration, each secondary user $i$ calculates its increase in throughput
if the best available channel (i.e., channel  $j_i^* = \mathop {\arg \max }\limits_{j \in {\mathcal{S}_a}} \: {p_{ij}}$)
is allocated. This increase in throughput can be calculated as follows:
\beqn
\label{Tlem1}
 \Delta {T_i} = T_i^a - T_i^b = \left[ {1 - \left( {1 - {p_{ij_i^*}}} \right)\prod\limits_{j \in \mathcal{S}_i } {(1 - {p_{ij}})} } \right] 
 \nonumber \\ 
 - \left[ {1 - \prod\limits_{j \in \mathcal{S}_i } {(1 - {p_{ij}})} } \right] 
= {p_{ij_i^*}} \prod\limits_{j \in \mathcal{S}_i } {(1 - {p_{ij}})}.  
\eeqn
It can be observed from (\ref{Tlem1}) that $\Delta {T_i}$ will quickly decrease over allocation iterations
because $\prod\limits_{j \in \mathcal{S}_i } {(1 - {p_{ij}})}$ tends to zero as the set $\mathcal{S}_i$ is expanded. We have the
following property for the resulting channel assignment due to Algorithm 1.

\begin{algorithm}[h]
\caption{\textsc{Non-Overlapped Channel Assignment}}
\label{mainalg}
\begin{algorithmic}[1]

\STATE Initialize the set of available channels  ${\mathcal{S}_a} := \left\{ {1,2, \ldots ,N} \right\}$ and $\mathcal{S}_i := \emptyset$ for $i=1, 2,\ldots , M$

\FOR{$i = 1$ to $M$}

\STATE $j_i^* = \mathop {\argmax }\limits_{j \in {\mathcal{S}_a}} \: {p_{ij}}$

\IF {$\mathcal{S}_{i} \neq 0$}

\STATE Find $\Delta {T_i} = T_i^a - T_i^b$,
where $T_i^a$ and $T_i^b$ is the throughputs after and before assigning channel $j_i^*$.

\ELSE

\STATE Find $\Delta {T_i} = p_{ij_i^*}$,

\ENDIF

\ENDFOR


\STATE ${i^*} = \argmax_i \Delta {T_i} $.

\STATE Assign channel $j_{i^*}^*$ to user $i^*$.

\STATE Update $\mathcal{S}_a = \mathcal{S}_a\backslash j_{i^*}^*$.

\STATE If $\mathcal{S}_a $ is empty, terminate the algorithm. Otherwise, return to step 2. 
\end{algorithmic}
\end{algorithm}

\vspace{0.2cm}
\noindent
\textbf{Proposition 1:} If we have $N >> M$, then the throughput achieved by any secondary user $i$ due to Algorithm 1 is very close
to the maximum value of 1.

\begin{proof}
This proposition can be proved by showing that if the number of channels is much larger than the number of
secondary users (i.e., $N >> M$) then each secondary user will be assigned a large number of channels.
Recall that Algorithm 1 assigns channels to a particular secondary user $i$ based on the increase-in-throughput metric $\Delta {T_i}$.
This property can be proved by observing that if a particular secondary user $i$ has been assigned a large
number of channels, its $\Delta {T_i}$ is very close to zero. Therefore, other secondary users who have been assigned a small
number of channels will have a good chance to receive more channels. As a result, all secondary users are assigned a large
number of channels if $N >> M$. According to (\ref{tput1}), throughput achieved by secondary user $i$ will reach its maximum
value of 1 if its number of assigned channels is sufficiently large. Hence, we have proved the proposition.
\end{proof}

In practice, we do not need a very large number of channels to achieve close-to-maximum throughput. In particular,
if each channel is available for secondary spectrum access with probability at least 0.8 then the throughput achieved
by a secondary user assigned three channels is not smaller than $1-(1-0.8)^3=0.992$, which is less than $1\%$ below the maximum
throughput. When the number of channel is not sufficiently large, we can potentially improve the system throughput by allowing
overlapped channel assignment. We develop overlapped channel assignment in the next section. 
After assigning channels using Algorithm 1 , i.e., a separate set at each secondary user is established, 
we calculate throughput of each secondary user by using (\ref{tput1}). Then,
the total throughput of the whole system can be calculated by summing throughput of all secondary users. 

\section{Overlapped Channel Assignment}
\label{over}

Overlapped channel assignment can improve the network throughput by exploiting the multiuser diversity gain.
In particular, a channel assigned to only one secondary user cannot be exploited if it is being used by a nearby
primary user. However, if a particular channel is assigned to several secondary users then it is more likely
that it can be exploited by at least one secondary user. However, when several
secondary users  attempt to access the same assigned channel, a MAC protocol is needed to resolve the access contention.
This MAC protocol incurs overhead that offsets the throughput gain due to the multiuser diversity. Hence, sophisticated
channel assignment algorithm is needed to balance the protocol overhead and throughput gain. 

\subsection{MAC Protocol}

\begin{figure}[!t]
\centering
\includegraphics[width=80mm]{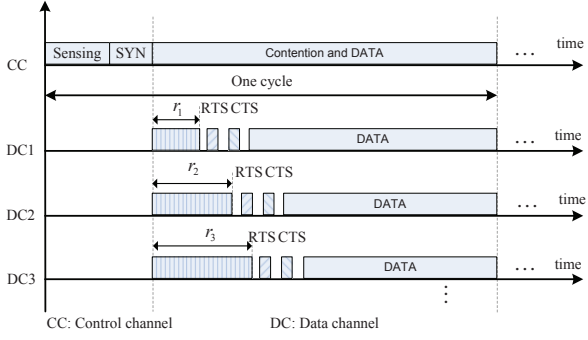}
\caption{Timing diagram for the proposed multi-channel MAC protocol.}
\label{MACoperation}
\end{figure}

\begin{algorithm}[h]
\caption{\textsc{Overlapped Channel Assignment}}
\label{mainalg}
\begin{algorithmic}[1]

\STATE Initialize the sets of allocated channels for all users  $\mathcal{S}_i := \emptyset$ for $i=1, 2,\ldots , M$ and $\delta_0$

\STATE Run Algorithm 1 to obtain non-overlapped channel assignment solution.

\STATE Let the group of channels shared by $l$ users be $\mathcal{G}_l$
and $U_j$ be the set of users sharing channel $j$ and set $U_j^\text{temp} := U_j, \: \forall j = 1, 2, \ldots, M$.

\STATE continue := 1; $h$ = 1; updoverhead := 0

\WHILE {$\text{continue}  = 1$}

\STATE Find the group of channels shared by $h$ users, $\mathcal{G}_{h}$

\FOR {$j=1$ to $\left|\mathcal{G}_{h}\right|$}

\FOR {$l=1$ to $M$}

\IF{$l \in U_j $}

\STATE $\Delta T_{l}^{h, \text{up}} (j) = 0$

\ELSE

\STATE User $l$ calculates $\Delta T_l^{h, \text{up}} (j)$ assuming channel $j$ is allocated to user $l$

\ENDIF

\ENDFOR

\STATE ${l^*_j} = \argmax_{l}  \Delta T_l^{h, \text{up}} (j) $.

\ENDFOR

\STATE ${j_{l^*}^*} = \argmax_j  \Delta T_{l^*_j}^{h, \text{up}} (j) $.

\IF{ $\Delta T_{l^*}^{h \text{up}} (j_{l^*}^*) \leq \epsilon$ and \text{updoverhead} = 1}

\STATE Set: \text{continue} := 0

\STATE Go to step 35

\ENDIF

\IF{ $\Delta T_{l^*}^{h, \text{up}} (j_{l^*}^*) > \epsilon$}

\STATE Temporarily assign channel $j_{l^*}^*$ to user $l^*$, i.e., 
 update $U_{j_{l^*}^*}^{\text{temp}} = U_{j_{l^*}^*}  \cup \left\{ l^* \right\} $;

\STATE Calculate $W$ and $\delta$ with  $U_{j_{l^*}^*}^{\text{temp}}$ by using methods in Sections \ref{ConWinCal} and \ref{Overcal}, respectively. 

\IF{ $\left| \delta - \delta_0 \right| > \epsilon_{\delta}$}

\STATE Set: \text{updoverhead} := 1

\STATE Return Step 7 using the updated $\delta_0 =\delta $

\ELSE 

\STATE Update  $U_{j_{l^*}^*} := U_{j_{l^*}^*}^{\text{temp}}$ 
 (i.e., assign channel $j_{l^*}^*$ to user $l^*$), calculate $W$ and $\delta_0$ with $U_{j_{l^*}^*}$, and
 update $\mathcal{G}_{h}$
 
 \STATE Update: \text{updoverhead} := 0

\ENDIF

\ENDIF

\STATE Return Step 7 

\STATE $h=h+1$

\ENDWHILE

\end{algorithmic}
\end{algorithm}

Let $\mathcal{S}_i$ be the separate set of
channels  assigned  for secondary user $i$ and $\mathcal{S}_{i}^{\sf com}$ be the set of  channels assigned for both users $i$ and other users.
Let denote $\mathcal{S}_i^{\sf tot} = \mathcal{S}_i \cup \mathcal{S}_i^{\sf com}$, which is the set of all channels assigned to user $i$. 
Assume that there is a control channel, which is always available and used for contention resolution of channel access.
We consider the following MAC protocol run by any particular secondary user $i$, which belongs the class of synchronized MAC
protocol \cite{R2}. After sensing assigned channels,
each user $i$ proceeds as follows. If there is at least one channel in $\mathcal{S}_i$ available, then user $i$ chooses
one of these available channels randomly for communication. If this is not the case, user $i$ will choose one available channel
 in $\mathcal{S}_i^{\sf com}$ randomly (if there is any channel in this set available) (for brevity we simply call \textit{users} instead of
 \textit{secondary users}  when there is no confusion). Then, it chooses
a random backoff value which is uniformly distributed in the interval $[0, W-1]$ (i.e.,  $W$ is the contention window) and 
starts decreasing its backoff counter while listening on the control channel.

If it overhears transmissions of RTS/CTS from any other users, it will freeze from decreasing its backoff counter until the control
channel is free again. As soon as a user's backoff counter reaches zero, its transmitter transmits an RTS message
containing a chosen channel to its receiver. If the receiver successfully receives the RTS, it will reply with
CTS and user $i$ starts its communication on the chosen channel for the remaining of the cycle.
In addition, by overhearing RTS/CTS messages of neighboring users, which convey information about the channels chosen for communications, other users compared these channels with their chosen ones. Any user who has his/her chosen channel 
coincides with the overheard channels quits the contention and waits until the next cycle. Otherwise, it will continue to decrease its backoff counter
before exchanging RTS/CTS messages. The MAC protocol is illustrated in Fig. ~\ref{MACoperation} where
sensing and synchronization phases are employed before the channel contention and transmission phase in each cycle. 
Note that the fundamental aspect that makes this MAC protocol different from
that proposed in \cite{Le11} is that in \cite{Le11} we assumed each winning user can use all available channels for
communications while at most one available channel can be exploited by hardware-constrained secondary users in this current paper.
Therefore, the channel assignment problem does not exist for the setting considered in  \cite{Le11}.

\subsection{Channel Assignment Algorithm}

We develop an overlapped channel assignment algorithm as follows.
First, we run Algorithm 1 to obtain the non-overlapped channel
assignment solution. Then, we start performing overlapped channel assignment
by allocating channels that have been assigned to a particular user
to other users. The MAC protocol overhead typically increases when a larger number
of secondary users compete for the same channel. Therefore, to achieve
the optimal tradeoff between overhead and the multiuser diversity gain,
only small number of users should share any channel.

We devise a greedy overlapped channel assignment algorithm using the increase-of-throughput metric similar to that
employed in Algorithm 1. 
However, calculation of this metric exactly turns out to be a complicated
task. Hence, we employ an estimate of the increase-of-throughput, which is derived in the following
to perform channel assignment assuming that the MAC protocol overhead  is $\delta<1$. 
In fact, $\delta$ depends on the outcome of the channel assignment algorithm (i.e., sets of
channels assigned to different users). Therefore, we will show
how to calculate $\delta$ and integrate it into this channel assignment algorithm later. 

Consider a case where channel $j$ is the common channel of users $i_1, i_2, \ldots, i_{\mathcal{MS}}$. Here,
$\mathcal{MS}$ is the number of users sharing this channel. We are interested in estimating
 the increase in throughput for a particular user $i$ if channel $j$ is assigned to this user.
 Indeed, this increase of throughput can be achieved because user $i$ may be able to exploit
 channel $j$ if this channel is not available or not used by other users $i_1, i_2, \ldots, i_{\mathcal{MS}}$.
 To estimate the increase of throughput, in the remaining of this paper we are only interested in
 a practical scenario where all $p_{ij}$ are close to 1 (e.g., at least 0.8). This would be a reasonable
 assumption given several recent measurements reveal that spectrum utilization of useful frequency bands
 is very low (e.g., less that $15\%$). Under this assumption,  
 we will show that the increase-of-throughput for user $i$ can be estimated as
\beqn \label{upith}
 \Delta T_{i}^{\mathcal{MS}, \text{up}} (j) =  
(1-1/\mathcal{MS})(1-\delta) p_{ij} \left( \prod_{h \in \mathcal{S}_i} \overline{p}_{ih} \right) \hspace{0.8cm} \label{Del1} \nonumber\\
\times  \left(1-\prod_{h \in \mathcal{S}_i^{\text{com}}}\overline{p}_{ih} \right)  \sum_{k=1}^{\mathcal{MS}} \left[ \overline{p}_{i_k j} \left(\prod_{q=1, q \neq k}^{\mathcal{MS}}p_{i_q j}  \right) \right]  \\
+(1-\delta) p_{ij}  \prod_{h \in \mathcal{S}_i} \overline{p}_{ih}    \prod_{h \in \mathcal{S}_i^{\text{com}}}\overline{p}_{ih}  \label{Del2} \hspace{3.2cm} \nonumber\\   
\times \prod_{q=1}^{\mathcal{MS}}p_{i_q j}   \prod_{q=1}^{\mathcal{MS}}\left(1-\prod_{h \in \mathcal{S}_{i_q}} \overline{p}_{i_q h}\right)   \hspace{0cm}\\
\hspace{0cm} + (1-1/\mathcal{MS})(1-\delta) p_{ij} \prod_{h \in \mathcal{S}_i} \overline{p}_{ih}  \left(1-\prod_{h \in \mathcal{S}_i^{\text{com}}}\overline{p}_{ih} \right)    \nonumber \\
\times  \prod_{q=1}^{\mathcal{MS}}p_{i_q j}  \prod_{q=1}^{\mathcal{MS}}\left(1-\prod_{h \in \mathcal{S}_{i_q}} \overline{p}_{i_q h}\right) \label{Del3}
\eeqn

This estimation is obtained by listing all possible scenarios/events where user $i$ can exploit channel $j$ to
increase its throughput. Because the user throughput is bounded by 1, we only count events that occur with
non-negligible probabilities. In particular, under the assumption that $p_{ij}$ are high (or $\overline{p}_{ij}$ are small)
we only count events whose probabilities have at most two such elements $\overline{p}_{ij}$ in the product. 
In addition, we can determine the increase of throughput for user $i$ by comparing its achievable throughput
before and after channel $j$ is assigned to it. It can be verified we have the following 
events for which the average increases of throughput are significant.
 
\begin{itemize}

\item Channel $j$ is available for all users $i$ and $i_q$, $q = 1, 2, \ldots, \mathcal{MS}$  except $i_k$ where $k = 1, 2, \ldots, \mathcal{MS}$. 
In addition, all channels in $S_i$ are not available and there is at least one channel in $\mathcal{S}_i^{\text{com}}$ available for user $i$. User $i$ can achieve a
maximum average throughput of $1-\delta$  by exploiting channel $j$, while its minimum average throughput before being assigned channel $i$
is at least $(1-\delta)/\mathcal{MS}$ (when user $i$ needs to share one available channel in $\mathcal{S}_i^{\text{com}}$ with $\mathcal{MS}$ 
other users). The increase of throughput for this case is $(1-1/\mathcal{MS})(1-\delta)$ and the upper-bound for the increase of
throughput of user $i$ is written in (\ref{Del1}).

\item Channel $j$ is available for user $i$ and all users $i_q$, $q = 1, 2, \ldots, \mathcal{MS}$ but each user $i_q$ uses other available channel in $\mathcal{S}_{i_q}$ for his/her transmission. Moreover, there is no channel in $\mathcal{S}_i^{\text{tot}}$ available. In this case, the increase of throughput for user $i$ is $1-\delta$ and the average increase of throughput of user $i$ is written in (\ref{Del2}).

\item Channel $j$ is available for user $i$ and all users $i_q$, $q = 1, 2, \ldots, \mathcal{MS}$ but each user $i_q$ uses other available channel in $\mathcal{S}_{i_q}$ for his/her transmission. Moreover, there is at least one channel in $\mathcal{S}_i^{\text{com}}$ available. In this case, the increase of throughput for user $i$ is upper-bounded by $(1-1/\mathcal{MS})(1-\delta)$ and the average increase of
throughput of user $i$ is written in (\ref{Del3}).

\end{itemize}

Detailed description of the algorithm is given in Algorithm 2. This algorithm has an outer and inter loops where the outer loop
increases the parameter $h$, which represents the maximum of users allowed to share any one particular channel (i.e., $\mathcal{MS}$ in
the above estimation of the increase of throughput) and the inner loop performs channel allocation for one particular value of $h=\mathcal{MS}$. 
In each assignment iteration of the inner loop, we assign one ``best'' channel $j$ to user $i$ that achieves maximum $\Delta T_{i}^{h,\text{up}} (j)$.
This assignment continues until the maximum $\Delta T_{i}^{h,\text{up}} (j)$ is less than a pre-determined number $\epsilon>0$.

\subsection{Calculation of Contention Window }
\label{ConWinCal}

We show how calculate contention window $W$ so that collision probabilities among contending secondary users is sufficiently small.
In fact, there is a trade-off between collision probabilities and the average overhead of the MAC protocol, which depends on $W$. In particular, larger values of $W$ reduce collision probabilities at the cost of higher protocol overhead and vice versa. Because there can be several collisions
during the contention phase each of which occurs if two or more secondary users randomly choose the same value of backoff time.
In addition, the probability of the first collision is largest because the number of contending users decreases
for successive potential collisions.

Let $\mathcal{P}_c$ be the probability of the first collision. In the following, we determine contention window $W$ by imposing a
 constrain $\mathcal{P}_c \leq \epsilon_P $ where $\epsilon_P $ controls the collision probability and overhead tradeoff. Let us calculate $\mathcal{P}_c$ as a function of $W$ assuming that there are $m$ secondary 
 users in the contention phase. Without loss of generality, assume that the random backoff times of $m$ secondary users are ordered as $r_1 \leq r_2 \leq \ldots \leq r_m$. The conditional probability of the first collision if there are $m$ secondary users in the contention stage can be written as
\beqn
\label{Pfirstc}
\mathcal{P}_c^{(m)} &=& \sum _{j=2}^{m} \Pr \left( j \: \text{users collide} \right) \nonumber\\
&=& \sum_{j=2}^m \sum_{i=0}^{W-2} C_m^j \left( \frac{1}{W}\right)^j \left( \frac{W-i-1}{W}\right)^{m-j}.
\eeqn
where each term in the double-sum represents the probability that $j$ users collide when they
choose the same backoff value equal to $i$. 
Hence, the probability of the first collision can be calculated as 
\beqn \label{pc}
\mathcal{P}_c = \sum_{m=2}^M \mathcal{P}_c^{(m)} \times \Pr\left\{m \: \text{ users contend}\right\},
\eeqn
where $\mathcal{P}_c^{(m)}$ is given in (\ref{Pfirstc}) and $\Pr\left\{m  \: \text{users contend}\right\}$
is the probability that $m$ secondary users join the contention phase. To compute $\mathcal{P}_c$, we now
derive $\Pr\left\{m  \: \text{users contend}\right\}$. It can be verified that 
secondary user $i$ joins contention if all channels in $\mathcal{S}_i$ are busy and there is at least one channel in $\mathcal{S}_i^{\sf com}$ available. The probability of this event can be written as
\beqn
\mathcal{P}_{\sf con}^{(i)} &=& \Pr \left\{ \text{all channels in} \: \mathcal{S}_i \: \text{ are busy}, \right. \nonumber\\
&& \left. \exists ! \: \text{some channels in} \:\mathcal{S}_i^{\sf com} \: \text{are available }\right\} \nonumber\\
&=&\left( \prod_{j \in \mathcal{S}_i} \overline{p}_{ij} \right) \left( 1- \prod_{j \in \mathcal{S}_i^{\sf com}} \overline{p}_{ij} \right).
\eeqn
The probability of the event that $m $ secondary users join the contention phase is 
\beqn \label{Pmusercon}
\Pr \left\{ m \: \text{users contend} \right\} = \sum_{n=1}^{C_M^m} \left( \prod_{i \in {\Lambda}_n} \mathcal{P}_{\sf con}^{(i)}\right) \hspace{1.5cm} \nonumber \\
\times \left( \prod_{j \in {\Lambda}_M \backslash {\Lambda}_n} \mathcal{\overline{P}}_{\sf con}^{(j)}\right) 
\eeqn
where ${\Lambda}_n$ is one particular set of $m$ users, ${\Lambda}_M$ is the set of all $M$ users ($\left\{ 1,2, \ldots , M \right\}$).
Substitute the result in (\ref{Pmusercon}) into (\ref{pc}), we can calculate $\mathcal{P}_c$. Finally,
we can determine $W$ as 
\beqn
W=  \min \left\{{W} \: \text{such that} \:  \mathcal{P}_c(W) \leq \epsilon_P \right\}
\eeqn
where for clarity we denote $\mathcal{P}_c(W)$, which is given in (\ref{pc}) as a function of $W$.

\subsection{Calculation of MAC Protocol Overhead}
\label{Overcal}

Let $r$ be the average value of the backoff value chosen by any secondary user.
Then, we have $r = (W-1)/2$ because the backoff counter value is uniformly chosen
in the interval $[0,W-1]$. As a result, average overhead can be calculated as follows:
\beqn
\delta\left(W\right) = \frac { \left[ W-1 \right]/2 \times \theta + t_{\sf RTS} + t_{\sf CTS} + 3 t_{\sf SIFS}} {\sf T_{\sf cycle}},
\eeqn
where $\theta$ is the time corresponding to one backoff unit;  $t_{\sf RTS}$,  $t_{\sf CTS}$, $t_{\sf SIFS}$
are the corresponding time of RTS, CTS and SIFS (i.e., short inter-frame space) messages; and $\sf T_{\sf cycle}$ is the cycle time. 
Here, we have assumed that the sensing and synchronization internals in each cycle are very short, which are, therefore, ignored in the overhead calculation.

\subsection{Update $\delta$ inside Algorithm 2}

Because the overhead $\delta$ depends on the channel assignment outcome, which is not known when we are 
running Algorithm 2. Therefore, in each allocation step we update $\delta$ based on the current channel assignment outcome.
Because $\delta$ does not change much in two consecutive allocation decisions, Algorithm 2 runs smoothly in practice.

\section{Numerical Results}
\label{Results}

\begin{figure}[!t]
\centering
\mbox{\subfigure[]{\includegraphics[width=1.7in]{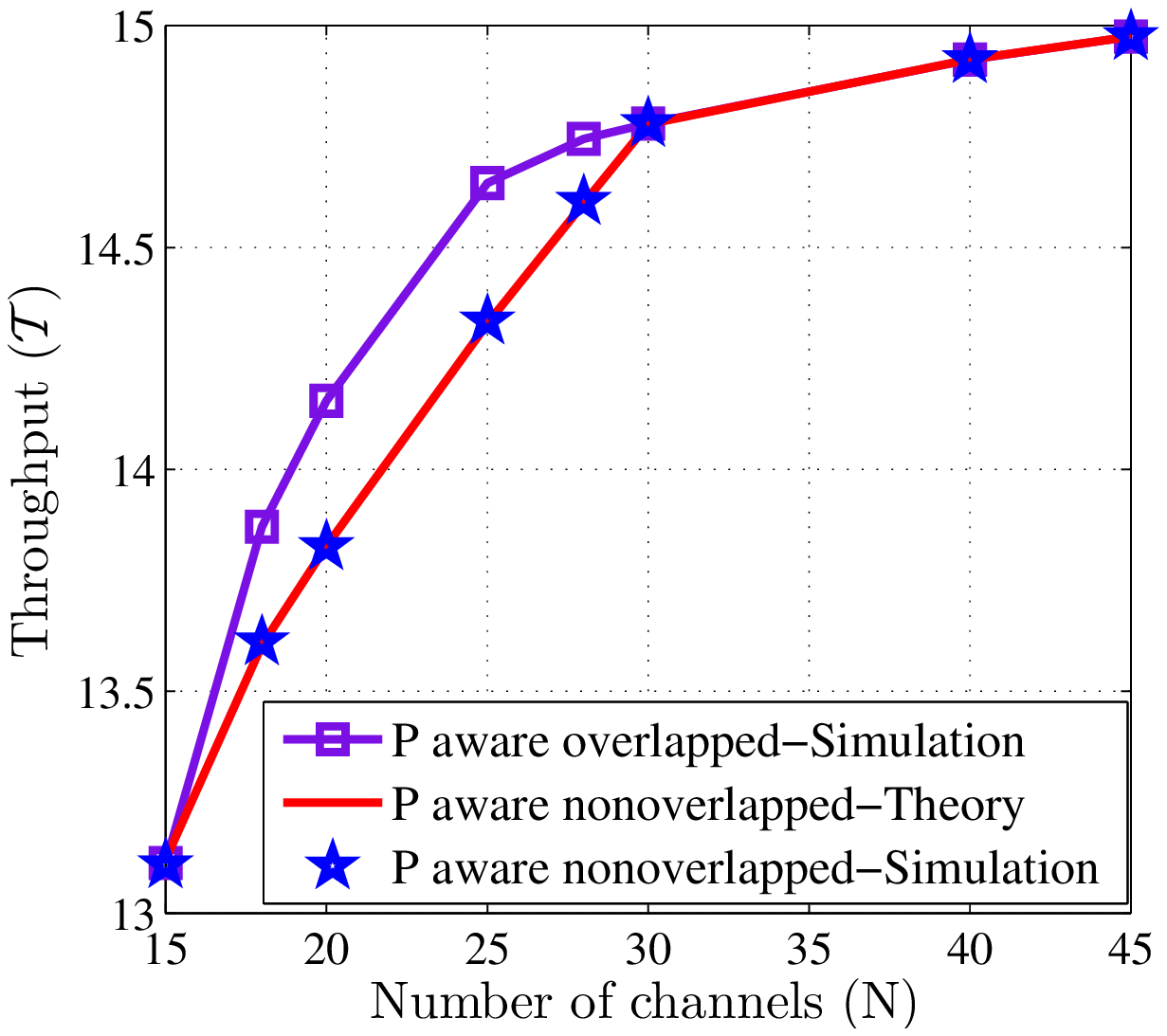} \label{het_T_M10_Aware_Blind1}}  
\subfigure[]{\includegraphics[width=1.7in]{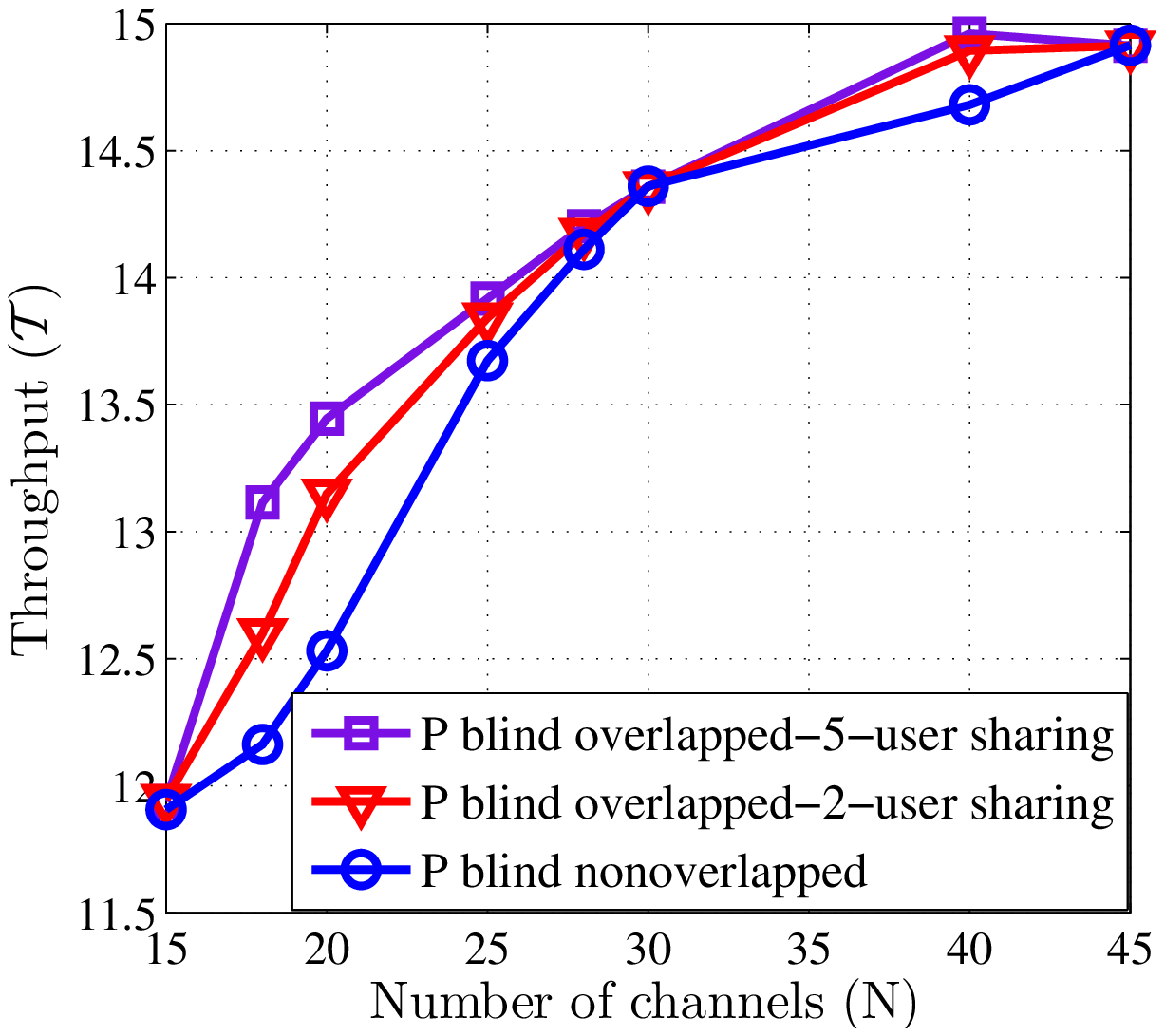} \label{het_T_M15_Aware_Blind1}} }
\caption{Throughput versus the number of channels (a) P-aware algorithms (Algs. 1 and 2) (b) P-blind algorithms}
\end{figure}

\begin{figure}[!t]
\centering
\mbox{\subfigure[]{\includegraphics[width=1.7in]{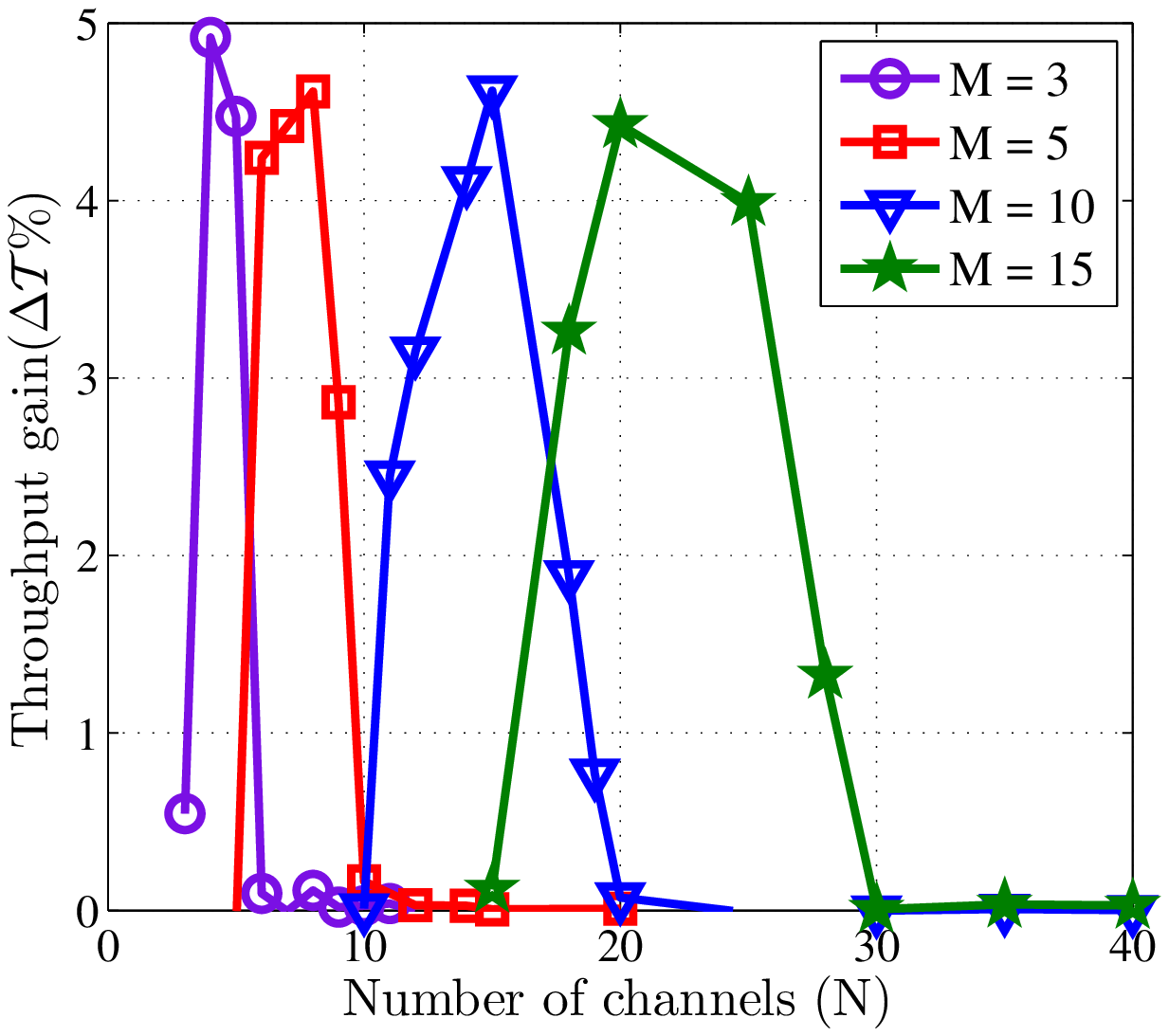} \label{het_G_M_351015}}  
\subfigure[]{\includegraphics[width=1.7in]{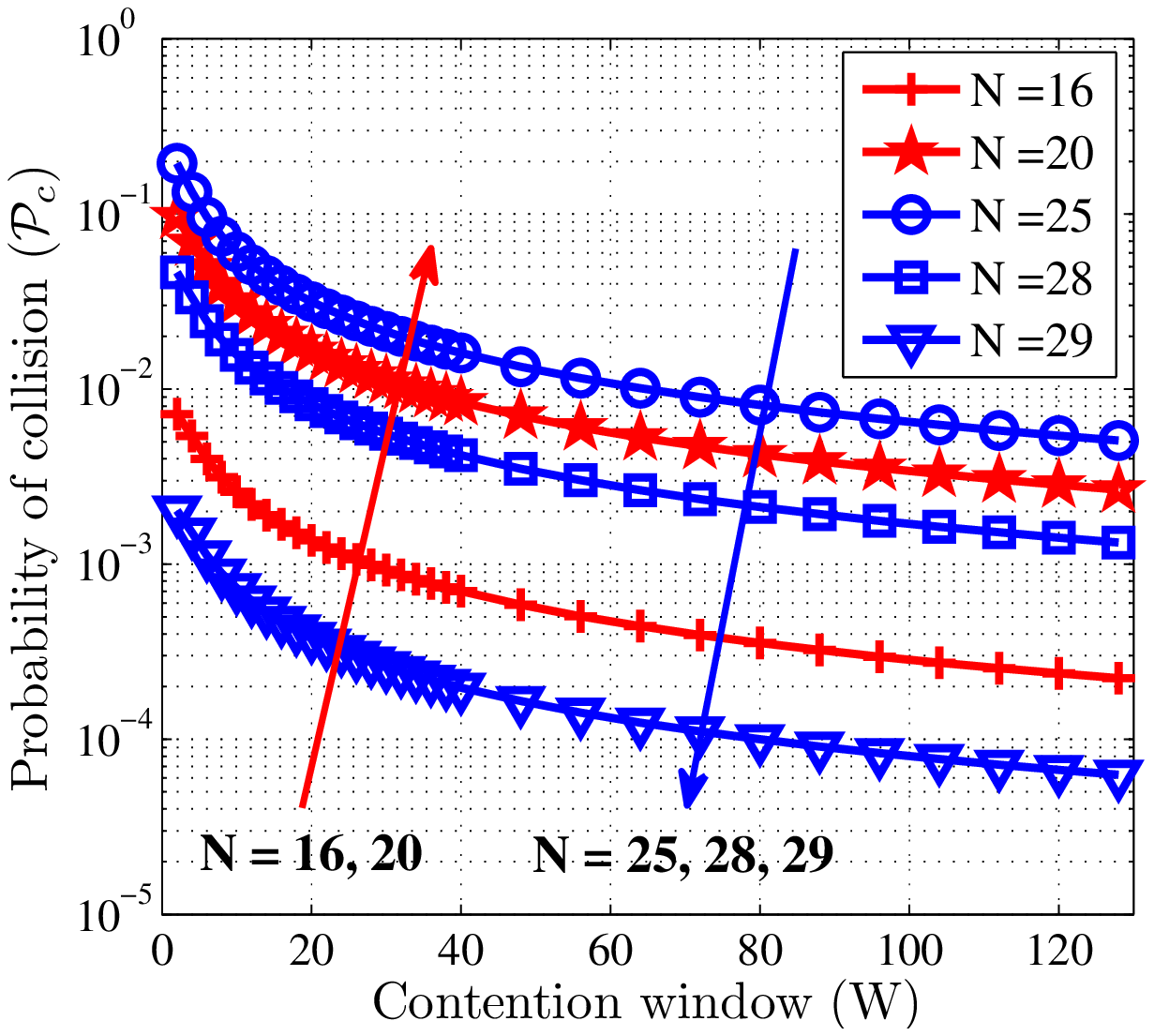} \label{het_P_M15_Aware_Blind}} }
\caption{(a) Throughput gain versus the number of channels (b) Collision probability versus the contention window (for M=15)}
\end{figure}

We present numerical results to illustrate the throughput performance of the proposed Algorithm 1 and 2. To 
test performance of both algorithms, the 
 probabilities $p_{i,j}$ are randomly realized in the interval [0.7, 0.9]. Other parameters are chosen as follows: cycle time 
 $T_{\sf cycle} = 3 ms$; $\theta = 20$ $\mu s$, $t_{\sf RTS}$ = 48$\mu s$, $t_{\sf CTS}$ = 40 $\mu s$,  $t_{\sf SIFS}$ = 15 $\mu s$, and
 target collision probability $\epsilon_P$ = 0.02. 
In Fig.~\ref{het_T_M10_Aware_Blind1},  we show total throughput $\mathcal{T}$ versus the number of channels $N$ for $M=15$ obtained by both Algs. 1 and 2 where each point is obtained by averaging the throughput over 30 different realizations of $p_{i,j}$. Throughput curves due to Algs. 1 and 2 are indicated as ``P-ware''  in the figures. In addition, for the comparison purposes, we also show throughput performance achieved by ``P-blind'' algorithms in Fig.~\ref{het_T_M15_Aware_Blind1}, which simply
allocate channels to users in a round-robin manner without considering particular values of $p_{i,j}$.

It can be seen that total throughput reaches the maximum value as the number of channels
becomes sufficiently large, which confirms the result stated in Proposition 1. In addition, Alg. 2 achieves significantly
larger throughput than Alg. 1 for lower values of $N$. This performance gain comes from the multiuser diversity gain,
which arises due to the spatial dependence of white spaces. 

As can be seen from Figs.~\ref{het_T_M10_Aware_Blind1} and \ref{het_T_M15_Aware_Blind1},
the proposed algorithms for both non-overlapped and overlapped cases outperform the round-robin channel assignment algorithms. For the non-overlapped case, Alg. 1 improves the total throughput significantly when comparing to the round-robin algorithm. For the overlapped case, we show throughput performance for the round-robin assignment algorithms when 2 and 5 users are allowed to share one channel (denoted as 2-user sharing and 5-user sharing in the figure). Although by allowing channel sharing among users, we can achieve larger throughput for the round-robin algorithm, they still perform worse compared to the proposed algorithms. We demonstrate the throughput gain due to Alg. 2 compared to Alg. 1 for different values
of $N$ and $M$ in Fig.~\ref{het_G_M_351015}. This figure shows that performance gains up to $5\%$ can be achieved by Alg. 2 compared to Alg. 1
when the number of channels is small or moderate. Moreover, we plot average probability of the first collision 
which is derived in Section IV.C versus contention window in Fig.~\ref{het_P_M15_Aware_Blind}. The outcomes
of Alg. 2 make the collision probability first increases then decreases with $N$. In fact, as $N$ is relatively small or large compared to $M$, the number of users sharing same channels is small, which leads to small collision probability.



\vspace{0.2cm}
\section{Conclusion}
\label{conclusion} 

We developed two channel assignment algorithms for throughput maximization in cognitive radio networks
with hardware-constrained secondary users. The first algorithm performed non-overlapped channel assignment 
for secondary users, which was shown to achieve optimality if the number of channels is sufficiently large. In the
secondary algorithm, we allowed overlapped channel assignments and designed a MAC protocol to resolve channel access
contention when different users attempt to exploit the same available channel. We validated our results via
numerical studies and demonstrated significant throughput gains of the overlapped channel assignment algorithm
compared to the non-overlapped counterpart in different network settings.

\bibliographystyle{IEEEtran}

\end{document}